\documentclass[final,3p,times]{elsarticle}
\usepackage{multirow,setspace,times,amssymb,amsmath,graphicx,color,rotating,subfigure,url}
\usepackage{lineno}
\usepackage{natbib}

\begin{document}

\begin{frontmatter}

\title{Price impact asymmetry of institutional trading in Chinese stock market}
\author[SB,RCE,RCSE]{Fei Ren\corref{cor}}
\cortext[cor]{Corresponding author. Address: 130 Meilong Road, P.O.
Box 114, School of Business, East China University of Science and
Technology, Shanghai 200237, China, Phone: +86 21 64253369, Fax: +86
21 64253152.}
\ead{fren@ecust.edu.cn} %
\author[SF]{Li-Xin Zhong}

\address[SB]{School of Business, East China University of Science and Technology, Shanghai 200237, China}
\address[SF]{School of Finance, Zhejiang University of Finance and Economics, Hangzhou 310018, China}
\address[RCE]{Research Center for Econophysics, East China University of Science and Technology, Shanghai 200237, China}
\address[RCSE]{Engineering Research Center of Process Systems Engineering (Ministry of Education), East China University of Science and Technology, Shanghai 200237, China}

\begin{abstract}
The asymmetric price impact between the institutional purchases and sales of 32 liquid stocks in Chinese stock markets in year 2003 is carefully studied. We analyze the price impact in both drawup and drawdown trends with consecutive positive and negative daily price changes, and test the dependence of the price impact asymmetry on the market condition. For most of the stocks institutional sales have a larger price impact than institutional purchases, and larger impact of institutional purchases only exists in few stocks with primarily increasing tendencies. We further study the mean return of trades surrounding institutional transactions, and find the asymmetric behavior also exists before and after institutional transactions. A new variable is proposed to investigate the order book structure, and it can partially explain the price impact of institutional transactions. A linear regression for the price impact of institutional transactions further confirms our finding that institutional sales primarily have a larger price impact than institutional purchases in the bearish year 2003.

\end{abstract}

\begin{keyword}
Econophysics; Price impact; Institutional trading; Market microstructure
\PACS 89.65.Gh, 89.75.Da, 05.45.Tp %
\end{keyword}

\end{frontmatter}

\section{Introduction}
\label{sec:Intro}

The study of large trades is of crucial importance for both market risk estimation and personal investments, since large trades generally have a strong impact on stock price and consequently increase investors' costs \cite{Ying-1966-Em,Karpoff-1987-JFQA,Wood-McInish-Ord-1985-JF,Gallant-Rossi-Tauchen-1992-RFS,Chan-Fong-2000-JFE,Lillo-Farmer-Mantegna-2003-Nature,Lim-Coggins-2005-QF,Naes-Skjeltorp-2006-JFinM,Zhou-2011-QF}. Meanwhile, growing evidence shows that large trades play a major role in trading in stock markets, which represent a large fraction of market's total trading volume \cite{Keim-Madhavan-1996-RFS,Jain-2003-JBF,Prino-Jarnecic-Lepone-2007-ABACUS,Gregoriou-2008-JES,Vaglica-Lillo-Mantegna-2010-NJP}. Therefore, there have been a variety of studies focusing on the price impact of large trades and the factors may influence their impact.

Large trades generally refer to those transactions executed with large number of shares. Kraus and Stoll firstly studied block trades traded in blocks of 10000 or more shares on the New York Stock Exchange, and found block purchases have a larger permanent impact than block sales \cite{Kraus-Stoll-1972-JF}. Keim and Madhavan studied the block trades in the U.S. markets, and found there exist significant differences between the temporary and permanent impacts of buyer-initiated and seller-initiated trades \cite{Keim-Madhavan-1996-RFS}. Gemmill collected the 20 largest customer purchases and sales for each share in the London Stock Exchange, and found that the impact of purchases is larger than the impact of sales, both temporarily and permanently \cite{Gemmill-1996-JF}.

An alternative way of investigating large trades is to study institutional trading, since institutions generally have large amounts of capital and their transactions have on average large number of shares. Chan and Lakonishok analyzed the price impact of the entire sequence of institutional trades in the New York and American Stock Exchanges, and found that the overall price impact after purchases and sales displays an intriguing asymmetry \cite{Chan-Lakonishok-1995-JF}. Chiyachantana {\em{et al.}} analyzed the institutional trading in 32 international stocks, and revealed that price impact asymmetry remarkably depends on the stock market condition: the price impact is much higher for institutional purchases in bull market, while institutional sales have a larger price impact in bear market  \cite{Chiyachantana-Jain-Jiang-Wood-2004-JF}.

Explanations appear in the literature to account for the price impact asymmetry between large purchases and sales. Saar used a trading model to investigate the behavior of institutional investors, and claimed that the information difference between buys and sells may cause the asymmetry \cite{Saar-2001-RFS}. However, this information explanation is hard to test from empirical measurement. Frino {\em{et al.}} and Gregoriou attributed  the price impact asymmetry to the bid-ask bias near open and close trades \cite{Gregoriou-2008-JES,Frino-Mollica-Walter-2003-XXX}. This price impact asymmetry can still be observed not using the opening or closing prices. Anderson measured the price impact using the transactions surrounding block trades, and related price effects to changes in order book depth \cite{Anderson-Cooper-Prevost-2006-FR}.

To the best of our knowledge, few studies have been conducted on the price impact asymmetry of institutional trading in the Chinese stock market. The main purpose of this paper is to test if the price impact asymmetry between institutional purchases and sales exists in the Chinese stock market. We will further attempt to study the dependence of the price impact asymmetry on the detailed condition of the Chinese stock market. Though it is still not clear what really causes the price impact asymmetry, a new variable introduced to facet the volumes and gaps of different price levels may offer a description of the order book structure surrounding institutional transactions. More remarkably, this variable can partially explain the price impact of institutional trading.

The rest of the paper is organized as follows. Section 2 introduces the data we analyzed and their summary statistics. Section 3 studies the price impact of institutional trading, and examines the asymmetry between the price impacts of purchases and sales. In Sections 4 and 5, we investigate the returns of trades and the order book structure related variable surrounding institutional transactions. Section 6 provides a regression model of explaining the price impact of institutional trading. Section 7 concludes.

\section{Data}

We use a sample of institutional trading data of 32 liquid stocks traded on the Shenzhen Stock Exchange (SZSE) in year 2003. The SZSE is one of the two stock exchanges in mainland China, and it has two separate markets including A-shares and B-shares. These 32 stocks analyzed in our study are all in the A-share market of SZSE, and their stock names are SDB (000001), VANKE-A (000002), CBG (000009), CSG (000012), KONKA GROUP (000016), KAIFA (000021), CMPD (000024), SEIC (000027), ZTE (000063), YANTIAN PORT (000088), SACL (000089), HYCC (000406), GPED (000429), SCPH (000488), GED (000539), FSL (000541), JMC (000550), WEIFU HIGH-TECH (000581), CHANGAN AUTOMOBILE (000625), HEBEI STEEL CORP. (000709), LN\&TS (000720), XINXING PIPES (000778), FAWCAR (000800), STSS (000825), CITIC GUOAN INFO. (000839), WULIANGYE (000858), ANSC (000898), TIK (000917), VALIN STEEL (000932), ZYTS (000956), and XSCE (000983).

The trading data are extracting from a database of order flows \cite{Gu-Chen-Zhou-2007-EPJB,Gu-Chen-Zhou-2008a-PA}. This database contains the orders of submission and cancelation of all the investors traded on the 32 stocks in year 2003. A double-auction mechanism is used during the opening time from 9:30 to 11:30 and 13:00 to 15:00, and transactions are automatically executed according to a price-time priority matching rule. In the database, each investor is endowed with a particular ID number. Therefore, we can obtain the trading records of each investor, including the trade price, the trading volume, and the transaction time. In addition, this database also provides a code identifying the investor type for each investor, i.e., institution or individual. Therefore, we can pick out the institution trading data, and analyze their price impact on stock prices.

Table~\ref{TB:stock:Summary} presents the summary statistics of the institution trading data of the 32 stocks. The stock codes are offered in the first column, and the industries they belong to are in the last column. The float capitalization $C_f$ (in units of million CNY (Chinese Yuan)), the number of institutional orders $N_o$, and the total size of institutional orders $S_o$ for each stock are provided in the second, third, and fourth columns respectively. Not exactly true, but in general those stocks with large number of institutional orders have large size of institutional orders. Table~\ref{TB:stock:Summary} also provides the corporate actions of the 32 stocks in year 2003, for instance cash dividend, bonus share, and rights issue. The main content of the corporate action includes the ex-dividend date, dividend payout ratio, and rights issue price.

\begin{table}[htp]
\begin{center}
  \caption{Summary statistics of the 32 stocks traded on the Shenzhen Stock Exchange in year 2003. We present the basic information about the 32 stocks, i.e., the stock code, the float capitalization $C_f$, the number of orders submitted by institutions $N_o$, the total size of orders submitted by institutions $S_o$, the corporate action, and the industrial sector the stock belonging to. The content of the corporate action includes the ex-dividend date, dividend payout ratio, and rights issue price.
}\label{TB:stock:Summary}
{\begin{tabular}{crrrll}
  \hline\hline
     Code & $C_f$ & $N_o$ & $S_o$ & Corporate action & Industrial sector\\
  \hline
 $000001$ & $12.0$ & $22820$ & $508501605$ & cash dividend, 0.15 CNY, September 29 & Financials\\%
 $000002$ & $6.1$  & $26498$ & $636849496$ & cash dividend, 0.2 CNY, May 23 & Real estate\\%
          &        &         &             & bonus share, 10:10, May 23   &\\
 $000009$ & $2.6$  & $3938$  & $82543467$  &   & Conglomerates\\%
 $000012$ & $0.8$  & $3328$  & $32130164$  & cash dividend, 0.15 CNY, June 18 & Metals $\&$ Nonmetals\\%
 $000016$ & $1.5$  & $7454$  & $89108621$  &   & Electronics\\%
 $000021$ & $2.2$  & $10298$ & $89128827$  & cash dividend, 0.1 CNY, July 8 & IT\\%
 $000024$ & $1.9$  & $9793$  & $107380350$ & cash dividend, 0.12 CNY, June 24 & Real estate\\%
          &        &         &             & rights issue, 3:10, 8.93 CNY, November 11   &\\
 $000027$ & $4.4$  & $33624$ & $580748895$ & cash dividend, 0.15 CNY, September 5 & Utilities\\%
 $000063$ & $4.7$  & $40795$ & $427661607$ & cash dividend, 0.2 CNY, May 23 & IT\\%
          &        &         &             & bonus share, 2:10, May 23   &\\
 $000066$ & $1.5$  & $6749$  & $47466245$  & cash dividend, 0.2 CNY, July 25 & IT\\%
 $000088$ & $3.0$  & $27677$ & $183665477$ & cash dividend, 0.1 CNY, July 28 & Transportation\\%
 $000089$ & $2.7$  & $22632$ & $415226048$ &  & Transportation\\%
 $000406$ & $2.2$  & $8241$  & $118973716$ & cash dividend, 0.3 CNY, August 14 & Petrochemicals\\%
 $000429$ & $1.4$  & $2910$  & $31087207$  & cash dividend, 0.1 CNY, June 13  & Transportation\\%
 $000488$ & $2.4$  & $13141$ & $96680682$  & cash dividend, 0.05 CNY, May 21 & Paper $\&$ Printing\\%
          &        &         &             & bonus share, 8:10, May 20   &\\
 $000539$ & $4.2$  & $26394$ & $129607010$ & cash dividend, 0.23 CNY, May 30 & Utilities\\%
 $000541$ & $1.8$  & $10124$ & $96995080$  & cash dividend, 0.42 CNY, June 23 & Machinery\\%
 $000550$ & $1.2$  & $24482$ & $222630276$ & cash dividend, 0.1 CNY, September 25 & Machinery\\%
 $000581$ & $2.5$  & $18188$ & $209812912$ & cash dividend, 0.2 CNY, August 1 & Machinery\\%
 $000625$ & $2.5$  & $20051$ & $273633251$ & cash dividend, 0.08 CNY, June 25 & Machinery\\%
 $000709$ & $3.0$  & $17929$ & $297208678$ & cash dividend, 0.25 CNY, June 19 & Metals $\&$ Nonmetals\\%
          &        &         &             & bonus share, 3:10, June 19   &\\
 $000720$ & $4.9$  & $3223$  & $12045770$  &  & Utilities\\%
 $000778$ & $2.5$  & $12987$ & $196427422$ & cash dividend, 0.5 CNY, May 22 & Metals $\&$ Nonmetals\\%
 $000800$ & $5.5$  & $42895$ & $1143849628$& cash dividend, 0.2 CNY, July 16 & Machinery\\%
 $000825$ & $3.0$  & $20302$ & $659599740$ & cash dividend, 0.05 CNY, August 22 & Metals $\&$ Nonmetals\\%
          &        &         &             & bonus share, 2:10, August 22   &\\
 $000839$ & $3.4$  & $20597$ & $211355321$ & cash dividend, 0.1 CNY, July 8 & Conglomerates\\%
 $000858$ & $4.3$  & $28747$ & $440712087$ & bonus share, 2:10, April 16 & Food $\&$ Beverage\\%
 $000898$ & $4.2$  & $32417$ & $1096933453$& cash dividend, 0.1 CNY, June 11 & Metals $\&$ Nonmetals\\%
 $000917$ & $1.3$  & $4204$  & $29235220$  & cash dividend, 0.1 CNY, June 27 & Media\\%
 $000932$ & $2.5$  & $32417$ & $1096933453$& cash dividend, 0.1 CNY, May 22 & Metals $\&$ Nonmetals\\%
 $000956$ & $3.0$  & $16373$ & $257413706$ & cash dividend, 0.1 CNY, June 3 & Petrochemicals\\%
          &        &         &             & rights issue, 2.5:10, 9.1 CNY, August 22 &\\
 $000983$ & $2.8$  & $22894$ & $290699030$ & cash dividend, 0.1 CNY, June 9 & Mining\\%
  \hline\hline
  \end{tabular}}
\end{center}
\end{table}

\section{Asymmetry price impact of institutional purchases and sales}

\subsection{Price impact}
In this section, we focus our attention on calculating the price impact of institutional trading. The price impact of block trades or institutional trading has been widely studied in a variety of stock markets \cite{Gemmill-1996-JF,Chan-Lakonishok-1995-JF,Holthausen-Leftwich-Mayer-1990-JF,Keim-Madhavan-1997-JFE,Conrad-Johnson-Wahal-2001-JF}. A popular way of measuring the price impact is to compare the average price of an executed order with an unperturbed price prior to the order. Following  Refs.~\cite{Chiyachantana-Jain-Jiang-Wood-2004-JF,Jones-Lipson-2001-JFE}, the price impact is calculated as the ratio of volume-weighted trade price ($P_{VWT}$) of the component trades in an order to the price at the time the order is released ($P_r$):
\begin{equation}
   PI = \ln P_{VWT} - \ln P_r
   \label{Eq:impact:buy}
\end{equation}
for institutional purchases, and
\begin{equation}
   PI = \ln P_r - \ln P_{VWT}
   \label{Eq:impact:sell}
\end{equation}
for institutional sales.
If we use $P_r$ to approximate the equilibrium market price prior to the transaction, $PI$ represents the total impact of the institutional trading \cite{Gregoriou-2008-JES,Anderson-Cooper-Prevost-2006-FR}. The total impact is known as the sum of the temporary and permanent impact in the literature study of block trades.

\subsection{Price impact asymmetry}
It has been revealed that the price impact is asymmetric between block purchases and sales, and this price impact asymmetry is essentially determined by the underlying market condition \cite{Gregoriou-2008-JES,Gemmill-1996-JF,Chiyachantana-Jain-Jiang-Wood-2004-JF,Saar-2001-RFS,Anderson-Cooper-Prevost-2006-FR,Keim-Madhavan-1997-JFE,Ellul-2006-JFE}. Chiyachantana {\em{et al.}} studied the institutional trading of international stocks from 37 countries, and found that purchases have a larger price impact than sales in the bull market, while sales have a larger price impact in the bear market \cite{Chiyachantana-Jain-Jiang-Wood-2004-JF}.

We have a database which contains the trading records of 32 stocks on the Chinese stock market in year 2003, and it was in the middle of a five-year bear market \citep{Zhou-Sornette-2004a-PA}. Our sample data do not cover two consecutive bull and bear market periods. To compensate for this problem, we use proxy methods to divide the price time series into drawups and drawdowns \cite{Maslov-Zhang-1999-PA,Johansen-Sornette-2001-JR,Rebonato-Gaspari-2006-QF}. A drawup (drawdown) is defined as a consecutive increase (decrease) in the daily price from a local minimum (maximum) to the next local maximum (minimum). Suppose we have a sequence of daily prices (specifically the closing prices), a drawup (drawdown) period is composed of several consecutive days with positive (negative) price changes. To make the drawup (drawdown) trend of the daily price more sustainable, we set a convention that a drawup (drawdown) is not stopped if the following drawdown (drawup) is not large enough. In our study, a drawup (drawdown) smaller than 30 percentage of the previous drawdown (drawup) or 3 times of the daily mean of the previous drawdown (drawup) is deemed not to have been interrupted. The choice of the convention does not change the results overall, only brings to some slight differences.

Since the share prices including the dividends and rights will cause abnormally large changes in daily prices, the share prices should be adjusted by excluding the dividends and rights when identifying the drawup and drawdown trends. In the Chinese stock market, cash dividend and bonus share are used as two common ways of paying dividends by listed companies to their shareholders. Table~\ref{TB:stock:Summary} describes the events of corporate actions including dividends and rights issues for the 32 stocks traded on the Shenzhen Stock Exchange in year 2003. For instance, stock 000001 paid a cash dividend of 0.15 CNY per share to its shareholders on September 29. Stock 000002 offered bonus shares in the ratio of 10:10 on May 23, which means the current shareholders could get ten free shares for ten shares they owns. Stock 000024 issued rights in the ratio of 3:10 on November 12, indicating for every 10 shares shareholders hold they could get another three at a deeply discounted price of 8.93 CNY.

\begin{figure}[h]
\centering
\includegraphics[width=8cm]{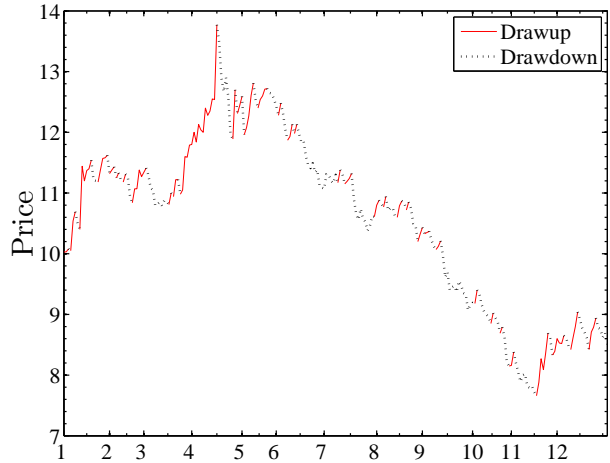}
\caption{(Color online) Daily price evolution of stock 000001 for 12 months in year 2003.}
\label{Fig:PE}
\end{figure}

After excluding the dividends and rights, we identify the drawup and drawdown trends in the daily prices for the 32 Chinese stocks. The daily price of stock 000001 evolved with the time for 12 months in year 2003 is illustrated in Figure~\ref{Fig:PE}, where the red solid curves represent the drawups and the black solid curves represent the drawdowns. It seems that most of the drawup and drawdown trends identified in our study are proper and sustainable.

In stead of computing the price impact of institutional trading in bull and bear market periods, we calculate the price impact in the drawup and drawdown periods. We combine the institutional transactions of 32 stocks, and find that the mean price impact $PI$ of sales is larger than that of purchases in both drawup and drawdown periods, as depicted in Table~\ref{TB:PriceImpact}. Our study confirms the previous finding that the sales have stronger effects than purchases in bear market \cite{Chiyachantana-Jain-Jiang-Wood-2004-JF}.

We further investigate the results for individual stocks. Interestingly, for some individual stocks, which primarily maintain increasing tendencies in year 2003, we observe the proof that the purchases have an impact larger than sales in drawup period. A variable $r$ is measured as the ratio of the number of days in drawup periods to the number of total trading days in 2003 to characterize the increasing tendency of a stock. According to the ratio $r$ we categorize the stocks into three groups, i.e., stocks with $r\geq0.55$, $r\leq0.45$, and $0.45< r < 0.55$. Among the ten stocks with relatively large ratio $r\geq 0.55$, four stocks (000002, 000024, 000063, and 000800) show that the purchases have an impact significantly larger than sales in drawup periods. Similar results are observed in another two stocks (000625 and 000983), but with relatively small $t$-Statistics. The larger impact of the institutional purchases also maintains in the drawdown periods: five stocks (000002, 000024, 000625, 000800, 000983) have significantly larger impact of purchases than that of sales, and two additional stocks (000063 and 000088) show similar behaviors with relatively small $t$-Statistics. For eight stocks with relatively small ratio $r\leq0.45$, seven stocks (000001, 000009, 000406, 000488, 000539, 000720, 000917) show that the impact of sales is larger than that of purchases in both drawup and drawdown periods, and six of them have significant $t$-statistics either in drawup or drawdown periods. The persistence of the larger impact of purchases or sales in both drawup and drawdown periods indicates that for those stocks with primarily increasing or decreasing tendencies the price impact asymmetry lasts over a relatively long period of time. For $0.45< r < 0.55$, the situation is more complex: in drawup periods, purchases have a significantly larger impact for stocks 000027, 000541, 000581, and 000778, and sales have a significantly larger impact for stocks 000089, 000825, 000839, 000858, and 000932; in drawdown periods, the purchases of stocks 000027, 000541, and 000778 show a significantly larger impact, and the sales of stocks 000016, 000825, 000839, 000858, and 000932 show a significantly larger impact.

\begin{table}[htp]
 \centering
 \caption{Mean price impact of institutional purchases and sales of the 32 stocks in the periods of price drawup and drawdown. The drawup (drawdown) is deemed not to have been interrupted if the following drawdown (drawup) is smaller than $30$ percentage of the previous drawup (drawdown) or 3 times of the daily mean of the previous drawup (drawdown). The stocks are categorized into three groups according to the ratio $r$, defined as the ratio of the number of days in drawup periods to the number of total trading days in 2003. The asterisks next to the t-statistic denote statistical significance at the $1\%$ level ($\ast\ast$) and $5\%$ level ($\ast$) for a two-tailed test that the mean price impact for purchases is different from that for sales.
 } \label{TB:PriceImpact}
\begin{tabular}{lcrrrcrrr}
  \hline\hline
    \multirow{3}*[2mm]{Stock codes} & \multirow{3}*[2mm]{r} & \multicolumn{3}{c}{Drawup} & & \multicolumn{3}{c}{Drawdown}\\
  \cline{3-5} \cline{7-9}
     & & Purchase & Sale & $t$-Statistic & & Purchase & Sale & $t$-Statistic\\
  \hline
  Total    &        & $7.97$  & $8.57$  &$-10.19^{\ast\ast}$ & & $8.41$  & $8.74$ & $-4.09^{\ast\ast}$\\
  \multicolumn{3}{l}{Stocks with ratio $r\geq0.55$}\\
  $000002$ & $0.60$ & $8.00$  & $6.72$  & $5.74^{\ast\ast}$  & & $7.99$  & $6.66$  & $4.28^{\ast\ast}$\\
  $000024$ & $0.55$ & $10.17$ & $9.12$  & $2.08^{\ast}$      & & $10.56$ & $8.51$  & $3.01^{\ast\ast}$\\
  $000063$ & $0.61$ & $7.49$  & $6.34$  & $3.97^{\ast\ast}$  & & $6.16$  & $5.72$  & $1.74$\\
  $000088$ & $0.63$ & $4.41$  & $7.19$  & $-11.09^{\ast\ast}$& & $7.36$  & $6.64$  & $1.54$\\
  $000550$ & $0.56$ & $8.07$  & $9.46$  & $-4.47^{\ast\ast}$ & & $8.09$  & $8.69$  & $-1.34$\\
  $000625$ & $0.55$ & $10.20$ & $9.77$  & $0.98$             & & $10.34$ & $9.00$  & $2.76^{\ast\ast}$\\
  $000709$ & $0.56$ & $6.78$  & $9.84$  & $-9.65^{\ast\ast}$ & & $8.45$  & $9.74$  & $-2.20^{\ast}$\\
  $000800$ & $0.61$ & $10.34$ & $9.01$  & $6.63^{\ast\ast}$  & & $9.66$  & $8.21$  & $5.19^{\ast\ast}$\\
  $000898$ & $0.65$ & $9.53$  & $10.33$ & $-2.55^{\ast}$     & & $9.69$  & $11.53$ & $-4.32^{\ast\ast}$\\
  $000983$ & $0.57$ & $8.81$  & $8.35$  & $1.74$             & & $11.04$ & $8.12$  & $6.33^{\ast\ast}$\\
  \multicolumn{3}{l}{Stocks with ratio $r\leq0.45$}\\
  $000001$ & $0.40$ & $6.96$  & $7.28$  & $-1.39$            & & $5.08$  & $9.52$  & $-15.41^{\ast\ast}$\\
  $000009$ & $0.35$ &$11.37$  &$12.99$  & $-1.91$            & & $10.48$ & $10.84$ & $-0.38$\\
  $000012$ & $0.37$ & $8.69$  & $7.32$  & $1.92$             & & $7.00$  & $6.61$  & $0.38$\\
  $000406$ & $0.44$ & $8.65$  & $9.58$  & $-2.05^{\ast}$     & & $8.90$  & $9.44$  & $-0.74$\\
  $000488$ & $0.42$ & $9.56$  & $9.63$  & $-0.11$            & & $9.88$  & $11.50$ & $-3.43^{\ast\ast}$\\
  $000539$ & $0.43$ & $4.56$  & $10.31$ & $-13.21^{\ast\ast}$& & $9.73$  & $12.36$ & $-2.98^{\ast\ast}$\\
  $000720$ & $0.42$ & $4.04$  & $7.69$  & $-7.42^{\ast\ast}$ & & $2.83$  & $9.25$  & $-9.12^{\ast\ast}$\\
  $000917$ & $0.41$ & $7.77$  & $8.73$  & $-1.37$            & & $8.75$  & $10.96$ & $-2.71^{\ast}$\\
  \multicolumn{3}{l}{Stocks with ratio $0.45<r<0.55$}\\
  $000016$ & $0.48$ & $10.58$ & $9.80$  & $1.41$             & & $9.19$  & $11.58$ & $-3.80^{\ast\ast}$\\
  $000021$ & $0.48$ & $8.88$  & $8.88$  & $0.01$             & & $7.49$  & $7.50$  & $-0.04$\\
  $000027$ & $0.52$ & $7.87$  & $6.89$  & $4.08^{\ast\ast}$  & & $7.42$  & $6.71$  & $2.55^{\ast}$\\
  $000066$ & $0.53$ & $9.39$  & $9.01$  & $0.64$             & & $7.94$  & $8.67$  & $-0.99$\\
  $000089$ & $0.47$ & $8.12$  & $9.22$  & $-3.41^{\ast\ast}$ & & $8.74$  & $8.54$  & $0.52$\\
  $000429$ & $0.46$ & $9.99$  & $11.19$ & $-1.37$            & & $13.16$ & $12.73$ & $0.31$\\
  $000541$ & $0.48$ & $8.49$  & $5.63$  & $7.67^{\ast\ast}$  & & $8.89$  & $6.97$  & $3.29^{\ast\ast}$\\
  $000581$ & $0.51$ & $11.06$ & $9.56$  & $3.50^{\ast\ast}$  & & $10.10$ & $9.83$  & $0.49$\\
  $000778$ & $0.54$ & $9.66$  & $5.81$  & $10.76^{\ast\ast}$ & & $8.89$  & $6.78$  & $4.29^{\ast\ast}$\\
  $000825$ & $0.53$ & $9.66$  & $10.99$ & $-4.28^{\ast\ast}$ & & $9.78$  & $12.38$ & $-5.53^{\ast\ast}$\\
  $000839$ & $0.52$ & $6.75$  & $7.80$  & $-4.21^{\ast\ast}$ & & $4.96$  & $7.44$  & $-7.20^{\ast\ast}$\\
  $000858$ & $0.49$ & $7.34$  & $10.54$ & $-13.78^{\ast\ast}$& & $6.98$  & $10.01$ & $-7.85^{\ast\ast}$\\
  $000932$ & $0.52$ & $10.37$ & $11.62$ & $-2.58^{\ast}$     & & $7.78$   & $11.33$ & $-6.46^{\ast\ast}$\\
  $000956$ & $0.53$ & $7.89$  & $7.97$  & $-0.27$            & & $8.09$   & $7.86$  & $0.65$\\
  \hline\hline
\end{tabular}
\end{table}

\section{Returns of trades surrounding institutional transactions}

In this section, we mainly investigate the returns caused by the trades surrounding the institutional transactions. We define the trade-by-trade return as the difference between the logarithmic prices of two consecutive trades
\begin{equation}
   R(t) = \ln P(t) - \ln P(t-1),
   \label{Eq:impact}
\end{equation}
where $P(t)$ is the price for the trade at event time $t$. The trades executed in a time range 60 seconds before and after the institutional transactions are collected as our experimental data. We aggregated the data from 32 stocks, and calculate the mean trade-by-trade return from -60 to 60 seconds around institutional transactions in an interval of 5 seconds. To further understand the dependence of the mean return of trades on the trading volumes of institutional transactions, the sample transactions are grouped into two subsets with equal size according to their trading volumes, i.e., subsets with small and large volume $V$.

Table~\ref{TB:PriceImpact:surrounding} provides the mean trade-by-trade return in the time range $T\in[-60,60]$ seconds centered around the institutional transactions in small and large volume subsets and the total institutional transactions, and $T=0$ denotes the time when the institutional transactions are executed. In the small $V$ subset, the mean return of trades before institutional purchases is positive, and displays a significant increase from -5 to $0^-$. This run-up in price consists with the earlier studies of the run-up prior to institutional purchases ~\cite{Gemmill-1996-JF,Anderson-Cooper-Prevost-2006-FR,Choe-Mclnish-Wood-1995-RQFA}. The mean trade-by-trade return surges from 0.17 in [-5,$0^-$] to 6.31 at $T=0$, and shows negative values indicating a clear price reversal following institutional purchases. This price reversal immediately after large trades has also been revealed in Refs.~\cite{Zawadowski-Kertesz-Andor-2004-PA,Mu-Zhou-Chen-Kertesz-2010-NJP}. The mean trade-by-trade return tends to be positive starting from the interval [25,30], and keeps relatively constant after that. The continuous positive impact on price may correspond to the permanent effect of institutional purchases ~\cite{Anderson-Cooper-Prevost-2006-FR}. Contrary to institutional purchases, the mean trade-by-trade return before institutional sales is mostly negative. The mean trade-by-trade return displays a significant decrease just before sales indicating a preceding run-down in price. The magnitude of its value surges to 6.53 at $T=0$, essentially larger than that before sales. A price reversal is also observed immediately after institutional sales, being followed with a negative impact permanently.

In the large $V$ subset, similar phenomena are observed for both institutional purchases and sales, including the run-up and run-down in the price prior to institutional purchases and sales, the price surge at $T=0$, the price reversal immediately after institutional transactions, and the permanent impact on the price later on. The absolute mean trade-by-trade return centered around institutional purchases in the large $V$ subset is significantly larger than that in the small $V$ subset. For institutional sales, the absolute mean trade-by-trade return in the large $V$ subset tends to be larger than that in the small $V$ subset starting from $T=0$, though they have comparable values before sales. This provides further evidence to prove the dependence of price impact of trades on their trading volumes \cite{Lillo-Farmer-Mantegna-2003-Nature,Zhou-2011-QF,Hasbrouck-1991-JF,Plerou-Gopikrishnan-Gabaix-Stanley-2002-PRE,Chordia-Subrahmanyam-2004-JFE}.

\begin{figure}[h]
\centering
\includegraphics[width=8cm]{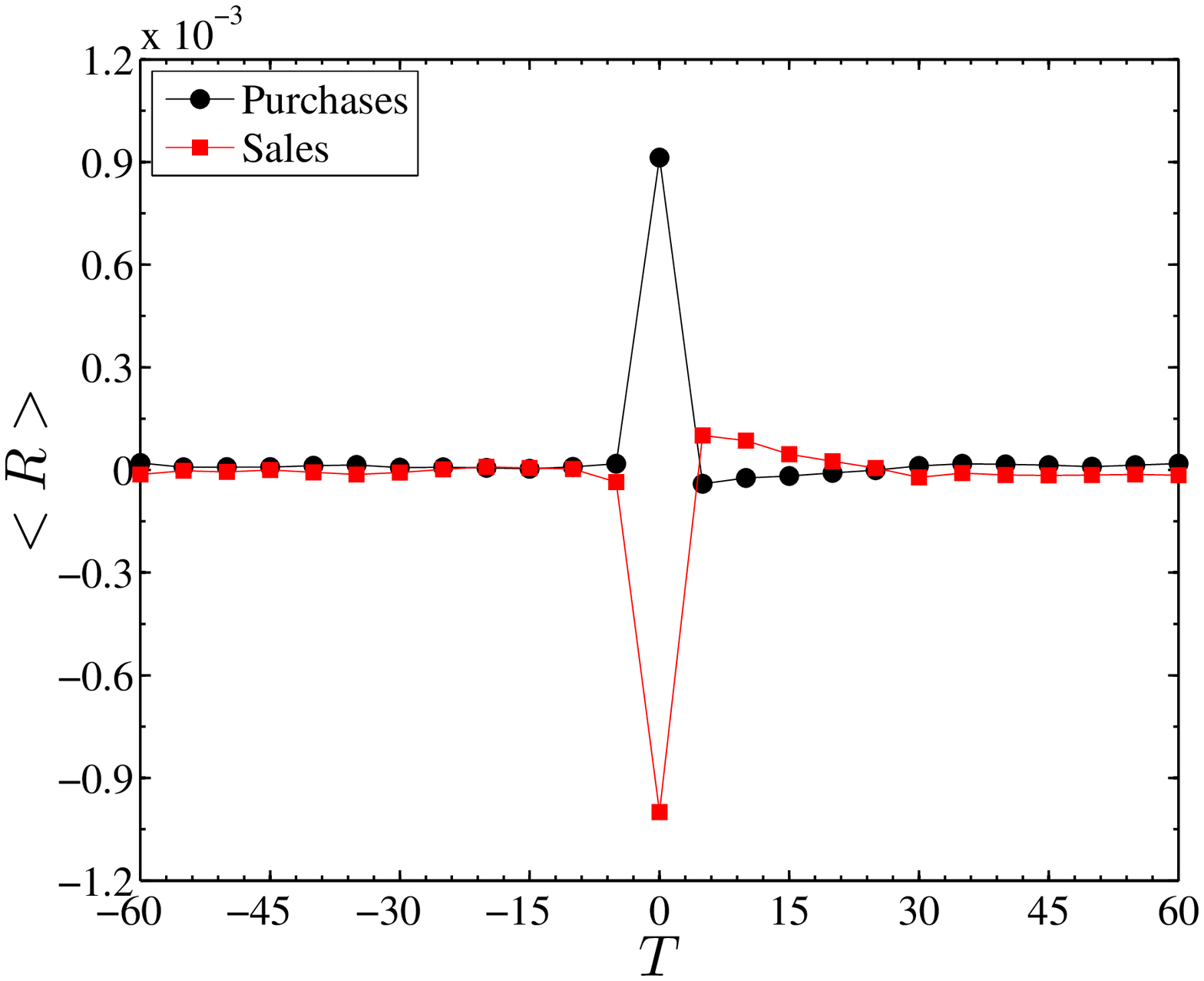}
\caption{(Color online) Mean trade-by-trade return in time range $[-60,60]$ seconds around total institutional purchases and sales.}
\label{Fig:PI:dynamics}
\end{figure}

The total transactions can be regarded as a combination of the transactions in small and large $V$ subsets, therefore, their mean trade-by-trade return shows a similar behavior basically consistent with that in both $V$ subsets. The mean trade-by-trade return is positive before institutional purchases and mostly negative before sales. A price run-up is observed prior to institutional purchases, and a run-down prior to sales. The absolute mean trade-by-trade return surges to a large value when institutional transactions are executed, and a price reversal occurs immediately after that. Later on a permanent impact persists from about 25 seconds after institutional transactions. To intuitively understand its dynamic behavior, we illustrate the mean trade-by-trade returns in time range $[-60,60]$ seconds around total institutional purchases and sales in Figure~\ref{Fig:PI:dynamics}.

\begin{table}[htp]
 \centering
 \caption{Mean trade-by-trade return $\langle R \rangle$ over the interval from -60 seconds to 60 seconds around institutional purchases and sales. The institutional transaction sequences are divided into two equal subsets based on their trading volumes. The mean returns around the institutional transactions in small and large volume subsets and the total institutional transactions are calculated respectively. The asterisks next to the t-statistic denote statistical significance at the $1\%$ level ($\ast\ast$) and $5\%$ level ($\ast$) for a two-tailed test that the absolute mean return for sales is different from that for purchases.
 } \label{TB:PriceImpact:surrounding}
\begin{tabular}{crrrcrrrcrrr}
  \hline\hline
    \multirow{3}*[2mm]{$T$} & \multicolumn{3}{c}{Small $V$} & & \multicolumn{3}{c}{Large $V$} & & \multicolumn{3}{c}{Total}\\
  \cline{2-4} \cline{6-8} \cline{10-12}
     & Purchases & Sales & $t$-Statistic & & Purchases & Sales & $t$-Statistic & & Purchases & Sales & $t$-Statistic\\
  \hline
  $-60\sim-55$ & $0.13$  & $-0.16$  & $-0.52$            & & $0.39$  & $-0.11$  & $4.87^{\ast\ast}$ & & $0.20$  & $-0.13$  & $1.86$\\
  $-55\sim-50$ & $0.05$  & $ 0.01$  & $1.21$             & & $0.17$  & $-0.05$  & $2.21^{\ast}$     & & $0.08$  & $-0.02$  & $1.65$\\
  $-50\sim-45$ & $0.01$  & $-0.05$  & $-0.80$            & & $0.30$  & $-0.06$  & $4.26^{\ast\ast}$ & & $0.08$  & $-0.05$  & $0.84$\\
  $-45\sim-40$ & $0.01$  & $ 0.08$  & $1.97^{\ast}$      & & $0.32$  & $-0.07$  & $4.37^{\ast\ast}$ & & $0.09$  & $ 0.01$  & $2.51^{\ast}$\\
  $-40\sim-35$ & $0.06$  & $-0.02$  & $0.70$             & & $0.36$  & $-0.10$  & $4.51^{\ast\ast}$ & & $0.12$  & $-0.07$  & $1.78$\\
  $-35\sim-30$ & $0.09$  & $-0.16$  & $-1.65$            & & $0.37$  & $-0.10$  & $4.74^{\ast\ast}$ & & $0.15$  & $-0.13$  & $0.66$\\
  $-30\sim-25$ & $0.04$  & $-0.03$  & $0.15$             & & $0.17$  & $-0.11$  & $1.10$            & & $0.07$  & $-0.07$  & $-0.14$\\
  $-25\sim-20$ & $0.04$  & $ 0.11$  & $3.59^{\ast\ast}$  & & $0.25$  & $-0.06$  & $3.28^{\ast\ast}$ & & $0.08$  & $ 0.02$  & $3.20^{\ast\ast}$\\
  $-20\sim-15$ & $0.04$  & $ 0.11$  & $3.99^{\ast\ast}$  & & $0.15$  & $ 0.07$  & $4.00^{\ast\ast}$ & & $0.06$  & $ 0.09$  & $5.15^{\ast\ast}$\\
  $-15\sim-10$ & $0.03$  & $ 0.03$  & $1.57$             & & $0.07$  & $ 0.09$  & $2.89^{\ast\ast}$ & & $0.04$  & $ 0.06$  & $3.35^{\ast\ast}$\\
   $-10\sim-5$ & $0.08$  & $-0.01$  & $2.68^{\ast\ast}$  & & $0.13$  & $ 0.08$  & $4.01^{\ast\ast}$ & & $0.09$  & $ 0.03$  & $5.08^{\ast\ast}$\\
   $-5\sim0^-$ & $0.17$  & $-0.39$  & $-13.69^{\ast\ast}$& & $0.24$  & $-0.25$  & $-0.11$           & & $0.18$  & $-0.35$  & $-11.00^{\ast\ast}$\\
        $0$    & $6.31$  & $-6.53$  & $-3.53^{\ast\ast}$ & & $12.61$ & $-12.89$ & $-3.30^{\ast\ast}$& & $9.13$  & $-10.00$ & $-16.17^{\ast\ast}$\\
    $0^+\sim5$ & $-0.20$ & $ 0.54$  & $19.58^{\ast\ast}$ & & $-1.75$ & $2.13$   & $6.98^{\ast\ast}$ & & $-0.40$ & $ 1.00$  & $34.79^{\ast\ast}$\\
     $5\sim10$ & $-0.13$ & $ 0.74$  & $21.10^{\ast\ast}$ & & $-0.82$ & $1.02$   & $3.33^{\ast\ast}$ & & $-0.23$ & $ 0.86$  & $24.93^{\ast\ast}$\\
    $10\sim15$ & $-0.14$ & $ 0.37$  & $6.67^{\ast\ast}$  & & $-0.38$ & $0.54$   & $2.78^{\ast\ast}$ & & $-0.18$ & $ 0.46$  & $9.97^{\ast\ast}$\\
    $15\sim20$ & $-0.08$ & $ 0.20$  & $3.33^{\ast\ast}$  & & $-0.13$ & $0.30$   & $2.92^{\ast\ast}$ & & $-0.09$ & $ 0.25$  & $5.57^{\ast\ast}$\\
    $20\sim25$ & $-0.04$ & $ 0.11$  & $1.88$             & & $ 0.12$ & $0.01$   & $2.18^{\ast}$     & & $-0.01$ & $ 0.06$  & $1.67$\\
    $25\sim30$ & $0.03$  & $-0.20$  & $-3.91^{\ast\ast}$ & & $ 0.44$ & $-0.23$  & $3.50^{\ast\ast}$ & & $0.12$  & $-0.22$  & $-2.88^{\ast\ast}$\\
    $30\sim35$ & $0.09$  & $ 0.08$  & $3.69^{\ast\ast}$  & & $ 0.51$ & $-0.24$  & $4.64^{\ast\ast}$ & & $0.18$  & $-0.09$  & $2.76^{\ast\ast}$\\
    $35\sim40$ & $0.09$  & $-0.09$  & $-0.07$            & & $ 0.43$ & $-0.20$  & $3.73^{\ast\ast}$ & & $0.17$  & $-0.15$  & $0.49$\\
    $40\sim45$ & $0.05$  & $-0.18$  & $-2.79^{\ast\ast}$ & & $ 0.43$ & $-0.15$  & $4.82^{\ast\ast}$ & & $0.14$  & $-0.16$  & $-0.59$\\
    $45\sim50$ & $0.02$  & $-0.12$  & $-2.15^{\ast}$     & & $ 0.31$ & $-0.17$  & $2.51^{\ast}$     & & $0.09$  & $-0.15$  & $-1.44$\\
    $50\sim55$ & $0.09$  & $-0.13$  & $-0.83$            & & $ 0.28$ & $-0.13$  & $2.75^{\ast\ast}$ & & $0.14$  & $-0.13$  & $0.42$\\
    $55\sim60$ & $0.11$  & $-0.06$  & $1.04$             & & $ 0.39$ & $-0.24$  & $2.58^{\ast\ast}$ & & $0.19$  & $-0.15$  & $0.95$\\
  \hline\hline
\end{tabular}
\end{table}

To accurately compare the mean trade-by-trade returns around institutional purchases and sales, we use two-sample T-tests to determine if the absolute mean trade-by-trade return for sales is different from that for purchases. Table~\ref{TB:PriceImpact:surrounding} provides the $t$-Statistics for the two-sample T-tests of the trade-by-trade returns from -60 to 60 seconds around the institutional transactions in small and large volume subsets and the total institutional transactions. Significant $t$-Statistics are observed in both positive and negative time intervals indicating asymmetric behaviors of mean trade-by-trade return before and after institutional transactions: For small $V$ subset the $t$-Statistic is statistically significant from the interval [-25,-20] to [45,50], wherein the absolute mean trade-by-trade return around institutional sales is significantly larger than that around purchases; For large $V$ subset the absolute mean trade-by-trade return before institutional purchases is significantly larger than that before sales starting from the interval [-60,-55], whereas it exhibits significantly larger values during and immediately after institutional sales than purchases; For total institutional transactions the absolute mean trade-by-trade return around sales is significantly larger than that around purchases from the interval [-20,-15] to [25,30], but with one exception [-10,-5]. In general, the absolute mean trade-by-trade return during and immediately after institutional sales is larger than that close to purchases.

\section{Structures of order books surrounding institutional transactions}

To better understand the price impact of institutional trading, we look into the order book structures before and after the institutional transactions. The limit order book refers to a list of pending limit orders submitted by market invertors, wherein limit orders are arranged in an accenting order according to their order prices. We investigate the order book structure by measuring the volumes on different price levels and the gaps between the prices of different levels, because these are two main variables characterizing the structure of the limit order book.

As a measure of the market liquidity, the order book depth has been faceted to explain the price impact of institutional purchases and sales \cite{Anderson-Cooper-Prevost-2006-FR}. The order book depth generally refers to the volumes that can be traded at or close to the current price. It has been also revealed that the gaps between different price levels in order book cause the large price movements \cite{Farmer-Gillemot-Lillo-Mike-Sen-2004-QF}. In this paper, we attempt to explain the price impact of institutional trading by considering both the volumes on different levels in the order book and the gaps between them, known as two variables characterizing the order book structure.

We generate the order book of a stock using the order flow data, and aggregate the data from the 32 Chinese stocks. Denote $a_i$ ($b_i$) as the price of a sell (buy) order on the level $i$, and $p_{mid}$ is the mid price between the best ask and bid. We introduce a composite variable $C$ to describe the order book structure on the opposite side of a market order, and assume that it follows the formula
\begin{equation}
   C = \sum_{i=1}^I ( a_i - p_{mid} ) v_i
   \label{Eq:impact:buy}
\end{equation}
for buy orders, and
\begin{equation}
   C = \sum_{i=1}^I ( p_{mid} - b_i ) v_i
   \label{Eq:impact:sell}
\end{equation}
for sell orders. It is measured as a sum of the product of the volume on the level $i$ and the gap between the price on the same level and the mid price. The sum is taken over the price levels $i=1,\cdots,I$, and $I$ is determined by the formula $\sum_{i=1}^I v_i =V$, where $V$ is the share volume fulfilled by a market order, i.e. the trading volume of a transaction.

The variable $C$ is not only regarded as a measure of the order book structure, more importantly, it can also be used to explain the price impact of institutional trading. Based upon the empirical findings that the price impact is attributed to both volumes and gaps of different price levels \cite{Anderson-Cooper-Prevost-2006-FR,Farmer-Gillemot-Lillo-Mike-Sen-2004-QF}, therefore, we simply assume that it obeys a product form as in Equations~(\ref{Eq:impact:buy}) and (\ref{Eq:impact:sell}). Furthermore, considering that only those executed shares have contributions to the price impact \cite{Lillo-Farmer-Mantegna-2003-Nature}, we take the sum of the volumes and gaps till the level $I$, which is determined by the effective trading volumes of a market order. This variable will be used as one of the explanatory variables of the price impact in next section.

Table~\ref{TB:Depth:surrounding} reports the mean variable $\langle C \rangle$ from -60 seconds to 60 seconds around the institutional purchases and sales in small and large volume subsets and the total institutional transactions. For small $V$ subset, $\langle C \rangle$ before and after institutional sales is generally larger than that around sales. An analysis of variance (ANOVA) is performed to test if $\langle C \rangle$ in different time intervals around institutional transactions are equal. Significant $F$-statistics are obtained for both institutional purchases and sales, moreover, $\langle C \rangle$ for institutional sales shows a maximum value at the time when sales are executed.

\begin{table}[htp]
 \centering
 \caption{Mean of variable $C$ over the interval from -60 seconds to 60 seconds around institutional purchases and sales. The institutional transaction sequences are divided into two equal subsets based on their trading volumes. The mean variable $\langle C \rangle$ around institutional transactions in small and large volume subsets and the total institutional transactions are calculated respectively. The asterisks next to the F-statistic denote the statistical significance at the $1\%$ level for a two-tailed test that $\langle C \rangle$ in different time intervals are equal.
 } \label{TB:Depth:surrounding}
\begin{tabular}{crrcrrcrr}
  \hline\hline
    \multirow{3}*[2mm]{$T$} & \multicolumn{2}{c}{Small $V$} & & \multicolumn{2}{c}{Large $V$} & & \multicolumn{2}{c}{Total}\\
  \cline{2-3} \cline{5-6} \cline{8-9}
     & Purchases & Sales & & Purchases & Sales & & Purchases & Sales \\
  \hline
  $-60\sim-55$ & $21.9$  & $37.8$  & & $62.5$  & $47.6$  & & $32.8$  & $43.0$\\
  $-55\sim-50$ & $22.2$  & $39.0$  & & $60.4$  & $45.3$  & & $32.3$  & $42.4$\\
  $-50\sim-45$ & $21.8$  & $31.5$  & & $64.6$  & $45.2$  & & $32.6$  & $38.7$\\
  $-45\sim-40$ & $21.1$  & $31.9$  & & $52.3$  & $48.0$  & & $28.5$  & $40.5$\\
  $-40\sim-35$ & $18.8$  & $36.4$  & & $54.9$  & $52.1$  & & $27.0$  & $44.6$\\
  $-35\sim-30$ & $15.9$  & $35.6$  & & $58.0$  & $45.4$  & & $25.0$  & $40.6$\\
  $-30\sim-25$ & $15.2$  & $33.3$  & & $58.7$  & $46.4$  & & $24.3$  & $40.2$\\
  $-25\sim-20$ & $13.8$  & $34.5$  & & $58.6$  & $47.4$  & & $22.7$  & $41.3$\\
  $-20\sim-15$ & $12.9$  & $30.8$  & & $51.6$  & $43.7$  & & $20.1$  & $37.4$\\
  $-15\sim-10$ & $12.4$  & $35.0$  & & $53.9$  & $44.5$  & & $19.7$  & $39.8$\\
   $-10\sim-5$ & $11.4$  & $28.9$  & & $58.0$  & $48.5$  & & $18.5$  & $37.2$\\
   $-5\sim0^-$ & $11.5$  & $28.9$  & & $63.1$  & $57.3$  & & $18.5$  & $37.0$\\
        $0$    & $19.0$  & $31.9$  & & $391.8$ & $405.8$ & & $189.0$ & $238.3$\\
    $0^+\sim5$ & $12.6$  & $30.3$  & & $70.4$  & $64.0$  & & $20.3$  & $40.2$\\
     $5\sim10$ & $11.5$  & $31.1$  & & $61.9$  & $48.6$  & & $19.1$  & $38.6$\\
    $10\sim15$ & $11.3$  & $33.0$  & & $60.5$  & $45.3$  & & $19.9$  & $39.3$\\
    $15\sim20$ & $10.5$  & $32.2$  & & $56.8$  & $49.3$  & & $19.2$  & $41.2$\\
    $20\sim25$ & $13.3$  & $36.3$  & & $60.8$  & $51.9$  & & $23.0$  & $44.6$\\
    $25\sim30$ & $14.2$  & $39.6$  & & $63.0$  & $47.7$  & & $24.8$  & $44.0$\\
    $30\sim35$ & $15.2$  & $41.2$  & & $62.3$  & $48.2$  & & $25.7$  & $44.9$\\
    $35\sim40$ & $15.1$  & $40.4$  & & $61.9$  & $49.7$  & & $26.1$  & $45.3$\\
    $40\sim45$ & $17.8$  & $42.0$  & & $56.0$  & $49.2$  & & $27.3$  & $46.0$\\
    $45\sim50$ & $19.1$  & $39.7$  & & $57.8$  & $44.0$  & & $29.4$  & $42.0$\\
    $50\sim55$ & $21.2$  & $34.8$  & & $53.3$  & $46.2$  & & $30.1$  & $40.9$\\
    $55\sim60$ & $21.7$  & $39.1$  & & $55.9$  & $50.0$  & & $31.5$  & $44.9$\\
    $F$-Statistic & $31.7^{\ast\ast}$  & $115.7^{\ast\ast}$  & & $1127.7^{\ast\ast}$ & $1658.0^{\ast\ast}$ & & $1342.0^{\ast\ast}$ & $1727.8^{\ast\ast}$\\
  \hline\hline
\end{tabular}
\end{table}

The mean variable $\langle C \rangle$ around the institutional transactions in large volume subset is worthy of careful study, since large institutional transactions may better represent large trades in stock markets. A trade with large volume is the fulfillment of a buy (sell) market order consuming large liquidity on sell (buy) side of the order book. By combining the results of $\langle C \rangle$ with the mean trade-by-trade return $\langle R \rangle$ in large $V$ subset, we might better understand the structure of the order book around large institutional transactions. For negative intervals, both $\langle C \rangle$ and $\langle R \rangle$ before institutional purchases are larger than those before sales. This indicates the price levels in order book are uniformly distributed before large institutional transactions. During the price reversals immediately after large institutional transactions, $\langle C \rangle$ after institutional purchases is larger than that after sales, while $\langle R \rangle$ after institutional purchases is smaller than that after sales. This may suggest that the sell side of the order book is thicker than buy side in a few seconds following large institutional transactions, i.e., large numbers of shares are complied on different price levels and the gaps between them are small. This phenomenon is intuitively reasonable as the investors are more incline to sell and less willing to buy in bear market. Following the price reversal, $\langle R \rangle$ after institutional purchases turns to be larger than that after sales, and the order book returns to be uniformly distributed.

\begin{figure}[h]
\centering
\includegraphics[width=8cm]{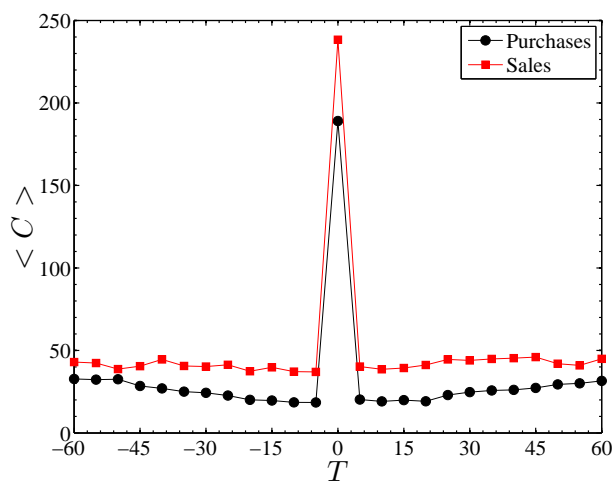}
\caption{(Color online) Mean variable $\langle C \rangle$ in time range $[-60,60]$ seconds centered around the total institutional transactions.}
\label{Fig:D:dynamics}
\end{figure}

The ANOVA obtains significant F-statistics for the mean variable $\langle C \rangle$ around the total institutional transactions, which indicates that $\langle C \rangle$ in different time intervals are not equal. For both institutional purchases and sales, $\langle C \rangle$ exhibits a maximum at the time when institutional transactions are executed as illustrated in Figure~\ref{Fig:D:dynamics}. This reminds us of the surge of the price impact at the time when institutional transactions are executed. Moreover, the curves have very similar shapes as those of the mean trade-by-trade return in the whole range of interval $T$. The mean variable $\langle C \rangle$ surrounding institutional sales is generally larger than that of purchases, and this can partially explain the larger impact of trades surrounding institutional sales except for the price reversals immediately after large institutional transactions. The similar patterns of $\langle C \rangle$ and $\langle R \rangle$ suggest a close relation between them, and the variable $C$ could be treated as one of the explanatory variables of the price impact.

\section{Modeling price impact of institutional transactions}

We model the price impact by taking into account the factors which may influence the price impact of the transactions performed by institutions. The explanatory variables include firm-specific factor, i.e. the capitalization, and order-specific factor, i.e., the order sign and the order size. The market related variables may also have significant effects on the price impact, such as the order book structure related variable $C$ and the volatility prior to institutional transactions. The dependent variable is the price impact $PI$ measured as the return of the volume weighted trading prices as defined in Equations~(\ref{Eq:impact:buy}) and ~(\ref{Eq:impact:sell}). The model we estimate follows a form:
\begin{equation}
   PI = \alpha + \beta_1 C_f + \beta_2 C + \beta_3 S + \beta_4 V_p + \epsilon.
   \label{Eq:impact:regression}
\end{equation}
The variable $C_f$ refers to the float capitalization of a firm. The variable $C$ introduced in previous section combines the contributions of the volumes on the different levels and the gaps between their prices and the mid price. A dummy variable $S$ is adopted to test the price impact asymmetry of institutional transactions. The variable $S$ takes the value 1 for sell trades, and 0 for buy trades. The price impact can also be affected by its previous volatilities, and $V_p$ is the mean absolute return within one minute before institutional transactions.

We aggregate the institutional transactions of 32 individual stocks, and normalize the variables, including $PI$, $C_f$, $C$, and $V_p$, by dividing their total standard deviations. In Table~\ref{TB:Regression}, the coefficients of variables estimated from the linear fit of Equation~(\ref{Eq:impact:regression}) are reported. We first make regression on the price impact of all institutional transactions done through market orders. The obtained $R$-Square is 0.143, and the fitted model could explain considerable portion of the variance in price impact. The coefficient $\beta_1$ is significantly negative at the $1\%$ level, showing a negative correlation between the firm's capitalization and the price impact. This is consistent with the commonsense observation that large capitalization companies have smaller price returns than those of small capitalization companies. The coefficient $\beta_2$ is positive and statistically significant, and it is further confirms our conjecture that institutional transactions with large variable $C$ have large price impact. The volatilities before institutional transactions also have a significant influence on the price impact of institutional transactions. More interestingly, the coefficient of dummy variable $\beta_3$ is positive and statistically significant at the $1\%$ level. This indicates that institutional sales have an impact on average larger than purchases, and it provides an additional proof of the asymmetric price impact between the institutional purchase and sales.

\begin{table}[htp]
 \centering
 \caption{Regression Coefficients of the price impact of all combined institutional transactions done through market orders using Equation~(\ref{Eq:impact:regression}). Separate regressions of the price impact for institutional purchases and sales are represented, and no dummy variable is assigned. Coefficients marked with asterisks are statistically significant at the $1\%$ level. The linear regression is also performed for the price impact of institutional transactions done through full-filled market orders, in which the market structure related variable $C$ is contributed from the total number of shares of market orders.
 } \label{TB:Regression}
\begin{tabular}{lrrrrrr}
  \hline\hline
  Variables  & All trades & $t$-Statistic & Purchases & $t$-Statistic & Sales & $t$-Statistic \\
  \hline
  \multicolumn{3}{l}{All market orders:}\\
  $\alpha$       & $0.444^{\ast\ast}$ & $115.7$ & $0.477^{\ast\ast}$ & $90.9$  & $0.452^{\ast\ast}$ & $87.6$\\
  $\beta_1$      & $-0.044^{\ast\ast}$& $-27.4$ & $-0.058^{\ast\ast}$& $-23.0$ & $-0.034^{\ast\ast}$& $-16.3$\\
  $\beta_2$      & $0.248^{\ast\ast}$ & $153.2$ & $0.229^{\ast\ast}$ & $106.6$ & $0.277^{\ast\ast}$ & $112.9$\\
  $\beta_3$      & $0.046^{\ast\ast}$ & $13.9$  &  &  &  &  \\
  $\beta_4$      & $0.263^{\ast\ast}$ & $162.4$ & $0.257^{\ast\ast}$ & $125.2$ & $0.275^{\ast\ast}$ & $103.8$\\
  $R$-Square     & $0.143$            &         & $0.134$            &         & $0.160$            &  \\
  \multicolumn{3}{l}{Full-filled market orders:}\\
  $\alpha$       & $0.400^{\ast\ast}$ & $88.4$  & $0.435^{\ast\ast}$ & $69.8$  & $0.433^{\ast\ast}$ & $70.4$\\
  $\beta_1$      & $-0.041^{\ast\ast}$& $-21.5$ & $-0.056^{\ast\ast}$& $-18.9$ & $-0.029^{\ast\ast}$& $-12.0$\\
  $\beta_2$      & $0.235^{\ast\ast}$ & $124.4$ & $0.223^{\ast\ast}$ & $88.5$  & $0.253^{\ast\ast}$ & $89.3$\\
  $\beta_3$      & $0.072^{\ast\ast}$ & $18.6$  &  &  &  & \\
  $\beta_4$     & $0.278^{\ast\ast}$ & $146.9$ & $0.272^{\ast\ast}$ & $115.6$ & $0.291^{\ast\ast}$ & $90.7$\\
  $R$-Square     & $0.146$            &         & $0.139$            &         & $0.157$            & \\
  \hline\hline
\end{tabular}
\end{table}

The linear regressions for the price impact of institutional purchases and sales are also performed separately. The regression form is similar as Equation~(\ref{Eq:impact:regression}), but the dummy variable $S$ is not assigned. The $R$-Square for institutional purchases and sales is 0.139 and 0.157 respectively. In general, significantly negative coefficient $\beta_1$ and significantly positive coefficients $\beta_2$ and $\beta_4$ are obtained for both institutional purchases and sales. Moreover, the coefficients $\beta_1$, $\beta_2$, and $\beta_4$ for institutional purchases are smaller than those for sales. This may further confirm our previous finding that institutional sales have larger price impact than purchases.

The price impact is measured by the volume-weighted trade prices of all the component trades of a market order submitted by institutions. However, the  variable $C$ only concerns the trading volumes executed immediately after the market order, which may be partial of the order size. One may argue that the it is not an appropriate factor of explaining price impact. Therefore, we further perform the regression for the price impact of institutional transactions done through full-filled market orders. In this case, the total number of shares of a market order are fulfilled immediately after its submission, therefore, the variable $C$ includes the contributions of the total shares composing a market order. As reported in the lower panels of Table~\ref{TB:Regression}, the regression coefficients for the price impact of the institutional transactions done through full-filled market orders show values similar to those of the institutional transactions of all market orders, and all of them are statistically significant.

\section{Conclusion}
\label{sec:Concl}
In this paper, we have studied the price impact of institutional trading of 32 liquid stocks in Chinese stock markets in year 2003, and find an asymmetric behavior between the institutional purchases and sales. Our results confirm the finding that institutional sales have a larger price impact than institutional purchases in bare market revealed in Ref.~\cite{Chiyachantana-Jain-Jiang-Wood-2004-JF}. To test the possible dependence of price impact asymmetry on market condition, we divide the price time series into drawup and drawdown trends. Only in four stocks with primarily increasing tendencies, we observe that institutional purchases have a larger impact than sales. This may provide a weak evidence of large price impact of institutional purchases in the market maintaining an upward trend.

We further investigate the mean trade-by-trade return surrounding institutional transactions, and observe an abundant phenomenon: a run-up (run-down) appears prior to institutional purchases (sales); the mean trade-by-trade return surges at the time when the institutional transaction occurs; a price reversal is observed immediately after institutional transactions; institutional trading has a permanent effect on stock price later after the price reversal. Remarkably, the asymmetric behavior is also observed before and after institutional transactions, showing different mean trade-by-trade returns surrounding institutional purchases and sales. A new variable $C$ is proposed to investigate the order book structure surrounding institutional transactions. This order book structure related variable shows a behavior very similar to that of the mean trade-by-trade return, and large (small) trade-by-trade returns are accompanied by large (small) $C$. The close relation between them may suggest that this variable could be regarded as one of the explanatory variables for price impact. Furthermore, the larger values of $C$ surrounding institutional sales may partially explain the larger impact of institutional sales on stock price.

A linear regression model is built to explain the price impact of institutional trading taking into account four explanatory variables, i.e., the capitalization of a firm, the order book structure related variable, a dummy variable denoting the order sign, and the previous volatility. The positive coefficient of the dummy variable provides further evidence that institutional sales have a larger price impact than purchases in the bearish year 2003. The coefficient of variable $C$ is positive and statistically significant, indicating that this variable is essential in explaining the price impact. Similar linear regressions of the price impact of institutional purchases and sales are performed separately, and the coefficient of $C$ for institutional sales is larger than that for purchases which further confirms the larger impact of sales on stock price.

\bigskip
{\textbf{Acknowledgments:}}

The authors thank Wei-Xing Zhou for helpful comments and discussions. This work was partially supported by the National Natural Science Foundation (Nos. 71131007 and 10905023), the Zhejiang Provincial Natural Science Foundation of China (Nos. Z6090130 and Y6110687), Humanities and Social Sciences Fund sponsored by Ministry of Education of the People's Republic of China (Nos. 09YJCZH042 and 10YJAZH137), and the Fundamental Research Funds for the Central Universities.

%\pagebreak
\bibliography{E:/papers/Auxiliary/Bibliography}

\begin{thebibliography}{10}
\expandafter\ifx\csname url\endcsname\relax
  \def\url#1{\texttt{#1}}\fi
\expandafter\ifx\csname urlprefix\endcsname\relax\def\urlprefix{URL }\fi

\bibitem{Ying-1966-Em}
C.~C. Ying, {Stock market prices and volumes of sales}, Econometrica 34 (1966)
  676--685.

\bibitem{Karpoff-1987-JFQA}
J.~M. Karpoff, {The relation between price changes and trading volume: A
  survey}, J. Financ. Quant. Anal. 22 (1987) 109--126.

\bibitem{Wood-McInish-Ord-1985-JF}
R.~A. Wood, T.~H. McInish, J.~K. Ord, {An investigation of transactions data
  for NYSE stocks}, J. Financ. 40 (1985) 723--739.

\bibitem{Gallant-Rossi-Tauchen-1992-RFS}
A.~R. Gallant, P.~E. Rossi, G.~Tauchen, {Stock prices and volume}, Rev. Financ.
  Stud. 5 (1992) 199--242.

\bibitem{Chan-Fong-2000-JFE}
K.~Chan, W.~M. Fong, {Trade size, order imbalance, and the volatility-volume
  relation}, J. Financ. Econ. 57 (2000) 247--273.

\bibitem{Lillo-Farmer-Mantegna-2003-Nature}
F.~Lillo, J.~D. Farmer, R.~Mantegna, {Master curve for price impact function},
  Nature 421 (2003) 129--130.

\bibitem{Lim-Coggins-2005-QF}
M.~Lim, R.~Coggins, {The immediate price impact of trades on the Australian
  Stock Exchange}, Quant. Financ. 5 (2005) 365--377.

\bibitem{Naes-Skjeltorp-2006-JFinM}
R.~N{\ae}s, J.~A. Skjeltorp, {Order book characteristics and the
  volume-volatility relation: Empirical evidence from a limit order market}, J.
  Financ. Markets 9 (2006) 408--432.

\bibitem{Zhou-2011-QF}
W.-X. Zhou, {Universal price impact functions of individual trades in an
  order-driven market}, Quant. Financ. 11 (2011) in press.

\bibitem{Keim-Madhavan-1996-RFS}
D.~B. Keim, A.~Madhavan, {The upstairs market for large-block transactions:
  analysis and measurement of price effects}, Rev. Financ. Stud. 9 (1996)
  1--36.

\bibitem{Jain-2003-JBF}
P.~K. Jain, {Discussion of ¡°Equity trading by institutional investors:
  Evidence on order submission strategies¡± by Naes and Skjeltorp}, J. Behav.
  Financ. 27 (2003) 1819--1821.

\bibitem{Prino-Jarnecic-Lepone-2007-ABACUS}
A.~Prino, E.~Jarnecic, A.~Lepone, {The determinants of the price impact of
  block trades: Further evidence}, ABACUS 43 (2007) 94--106.

\bibitem{Gregoriou-2008-JES}
A.~Gregoriou, {The asymmetry of the price impact of block trades and the
  bid-ask spread Evidence from the London Stock Exchange}, J. Econ. Surveys 35
  (2008) 191--199.

\bibitem{Vaglica-Lillo-Mantegna-2010-NJP}
G.~Vaglica, F.~Lillo, R.~N. Mantegna, {Statistical identification with hidden
  Markov models of large order splitting strategies in an equity market}, New
  J. Phys. 12 (2010) 075031.

\bibitem{Kraus-Stoll-1972-JF}
A.~Kraus, H.~R. Stoll, {Price impact of block trading on the New York Stock
  Exchange}, J. Financ. 27 (1972) 569--588.

\bibitem{Gemmill-1996-JF}
G.~Gemmill, {Transparency and liquidity: A study of block trades in the London
  Stock Exchange under different publication rules}, J. Financ. 51 (1996)
  1765--1790.

\bibitem{Chan-Lakonishok-1995-JF}
K.~C. Chan, J.~Lakonishok, {The behavior of stock prices around institutional
  trades}, J. Financ. 50 (1995) 1147--1174.

\bibitem{Chiyachantana-Jain-Jiang-Wood-2004-JF}
C.~N. Chiyachantana, P.~K. Jain, C.~Jiang, R.~A. Wood, {International evidence
  on institutional trading behavior and price impact}, J. Financ. 59 (2004)
  869--898.

\bibitem{Saar-2001-RFS}
G.~Saar, {Price impact asymmetry of block trades: An institutional trading
  explanation}, Rev. Financ. Stud. 14 (2001) 1153--1181.

\bibitem{Frino-Mollica-Walter-2003-XXX}
A.~Frino, V.~Mollica, T.~Walter, {Asymmetric price behaviour surrounding block
  trades: A market microstructure explanation} (2003).

\bibitem{Anderson-Cooper-Prevost-2006-FR}
H.~D. Anderson, S.~Cooper, A.~K. Prevost, {Block trade price asymmetry and
  changes in depth: Evidence from the Australian stock exchange}, Financ. Rev.
  41 (2006) 247--271.

\bibitem{Gu-Chen-Zhou-2007-EPJB}
G.-F. Gu, W.~Chen, W.-X. Zhou, {Quantifying bid-ask spreads in the Chinese
  stock market using limit-order book data: Intraday pattern, probability
  distribution, long memory, and multifractal nature}, Eur. Phys. J. B 57
  (2007) 81--87.

\bibitem{Gu-Chen-Zhou-2008a-PA}
G.-F. Gu, W.~Chen, W.-X. Zhou, {Empirical distributions of Chinese stock
  returns at different microscopic timescales}, Physica A 387 (2008) 495--502.

\bibitem{Holthausen-Leftwich-Mayer-1990-JF}
R.~W. Holthausen, R.~W. Leftwich, D.~Mayers, {Large block transactions, speed
  of response, and temporary and permanent stock price effects}, J. Financ. 26
  (1990) 71--95.

\bibitem{Keim-Madhavan-1997-JFE}
D.~B. Keim, A.~Madhavan, {Transaction costs and investment style: An
  interexchange analysis of institutional equity trades}, J. Financ. Econ. 46
  (1997) 265--292.

\bibitem{Conrad-Johnson-Wahal-2001-JF}
K.~M. Conrad, J. S.and~Johnson, S.~Wahal, {Institutional trading and soft
  dollars}, J. Financ. 56 (2001) 397--417.

\bibitem{Jones-Lipson-2001-JFE}
C.~M. Jones, M.~L. Lipson, {Sixteenths: Direct evidence on institutional
  execution costs}, J. Financ. Econ. 59 (2001) 253--278.

\bibitem{Ellul-2006-JFE}
A.~Ellul, {Ripples through markets: Inter-market impacts generated by large
  trades}, J. Financ. Econ. 82 (2006) 173--196.

\bibitem{Zhou-Sornette-2004a-PA}
W.-X. Zhou, D.~Sornette, {Antibubble and prediction of China's stock market and
  real-estate}, Physica A 337 (2004) 243--268.

\bibitem{Maslov-Zhang-1999-PA}
S.~Maslov, Y.~C. Zhang, {Probability distribution of drawdowns in risky
  investments}, Physica A 262 (1999) 232--241.

\bibitem{Johansen-Sornette-2001-JR}
A.~Johansen, D.~Sornette, {Large stock market price drawdowns are outliers}, J.
  Risk 4~(2) (2001) 69--110.

\bibitem{Rebonato-Gaspari-2006-QF}
R.~Rebonato, V.~Gaspari, {Analysis of drawdowns and drawups in the US\$
  interest-rate market}, Quant. Financ. 6 (2006) 297--326.

\bibitem{Choe-Mclnish-Wood-1995-RQFA}
H.~Choe, T.~Mclnish, R.~Wood, {Block versus nonblock trading patterns}, Review
  of Quantitative Finance and Accounting 5 (1995) 355--363.

\bibitem{Zawadowski-Kertesz-Andor-2004-PA}
A.~G. Zawadowski, J.~Kert{\'e}sz, G.~Ador, {Large price changes on small
  scales}, Physica A 344 (2004) 221--226.

\bibitem{Mu-Zhou-Chen-Kertesz-2010-NJP}
G.-H. Mu, W.-X. Zhou, W.~Chen, J.~Kert{\'e}sz, {Order flow dynamics around
  extreme price changes on an emerging stock market}, New J. Phys. 12 (2010)
  075037.

\bibitem{Hasbrouck-1991-JF}
J.~Hasbrouck, {Measuring the information content of stock trades}, J. Financ.
  46 (1991) 179--207.

\bibitem{Plerou-Gopikrishnan-Gabaix-Stanley-2002-PRE}
V.~Plerou, P.~Gopikrishnan, X.~Gabaix, H.~E. Stanley, {Quantifying stock-price
  response to demand fluctuations}, Phys. Rev. E 66 (2002) 027104.

\bibitem{Chordia-Subrahmanyam-2004-JFE}
T.~Chordia, A.~Subrahmanyam, {Order imbalance and individual stock returns:
  Theory and evidence}, J. Financ. Econ. 72 (2004) 485--518.

\bibitem{Farmer-Gillemot-Lillo-Mike-Sen-2004-QF}
J.~D. Farmer, L.~Gillemot, F.~Lillo, S.~Mike, A.~Sen, {What really causes large
  price changes?}, Quant. Financ. 4 (2004) 383--397.

\end{thebibliography}
%\bibliography{E:/paper/bibfile/Bibliography}

\end{document}